\documentclass[pre,floatfix,10pt]{revtex4} 
\usepackage{amssymb,euscript,latexsym,amsmath}
\usepackage{graphicx}
\usepackage{pgfplots}
\usepackage{amsmath}
\usepackage{appendix}
\usepackage{tikz}
\usepackage{soul}
\usepackage[normalem]{ulem}
\usepackage[customcolors,shade]{hf-tikz}
\usepackage{amsmath}
\begin{document}
\title{
%Rainbows and boat wakes: Unveiling a Common  Principle
%\\o
%\\
Common principles behind rainbows and boat wakes}

\author{Eduardo A. Jagla}
\email{eduardo.jagla@ib.edu.ar}
\affiliation{Comisión Nacional de Energía Atómica and Instituto Balseiro
CNEA, CONICET, UNCUYO
Av. E. Bustillo 9500 S. C. de Bariloche, Argentina}
\author{Alberto G. Rojo}
\email{rojo@oakland.edu}
\affiliation{Department of Physics, Oakland University, Rochester, MI 48309}

\begin{abstract}

Rainbows and boat wakes may seem unrelated, but they share deep mathematical connections through ray folding, caustics, and Airy interference. This paper explores these principles, which are also relevant for explaining phenomena such as shimmering effects on the bottom of pools and twinkling stars. By revisiting Airy’s theories on wavefronts and caustics, we demonstrate their applications not only in optics and for water waves but also in quantum wave packets. Using concepts from undergraduate physics, we highlight the universal patterns that unify these diverse phenomena.

%
%% NEW
%\textcolor{blue}{We find very good agreement with the observed velocities of metors from many %different comets}
%%

%%
%% NEW
%\textcolor{blue}{text}
%%
\end{abstract}
\maketitle
%
%\newpage
\begin{flushright}
{``Caminante no hay camino,

{sino} estelas en la mar.''

({\em{Traveler, there is no road;

only a ship's wake on the sea}})

Antonio Machado\cite{Machado}
}
\end{flushright}

\begin{flushright}
{``Why are there so many songs about rainbows

And what's on the other side"

Paul Williams\cite{Williams}
}
\end{flushright}
\section{Introduction}

Rainbows and boat wakes are two examples of wave phenomena, the first related to electromagnetism
and the second involving gravity waves on a liquid surface. 
Despite this shared property, the two phenomena may seem unrelated at first glance.
%Beyond this overall similarity, the
%two seem unrelated at first glance. 
However, a deep mathematical and physical relation exists between them. 
Although the physics of rainbows and boat wakes is well understood individually\cite{Crawford0,Boyer}, the pedagogical value of this work lies in emphasizing the unifying principles that connect them.

These principles include ray folding, caustics, and Airy interference, which
also play a central role in other everyday phenomena, such as
the shimmering patterns that can be observed on the bottom of a swimming pool, the twinkling of stars, and the reflections 
of light inside a cup of tea.

\begin{figure}[h]
\includegraphics[width=8cm,clip=true]{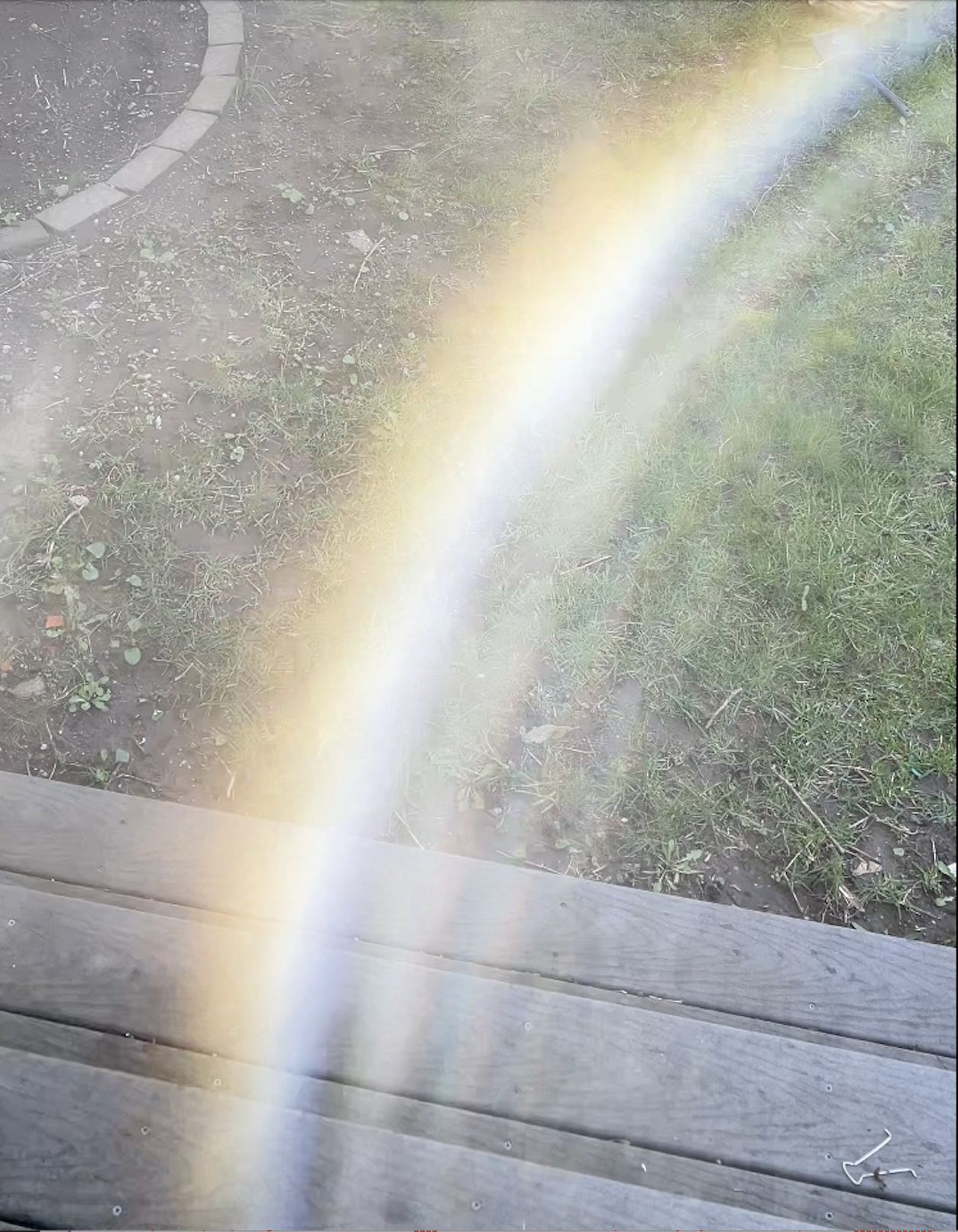}
\caption{(Color online) Supernumerary arcs generated on a sunny day from drops produced by a flower mister.  Photograph by one of the authors (A.G.R.)
%(Permission pending from the source)
}
\label{SupernumeraryPhoto}
\end{figure}

The modern theory of the rainbow is due to George Biddle Airy\cite{Airy}, who improved upon the Cartesian-Newtonian ray optics theory. Specifically, ray optics failed to explain how the water droplet size influences the colors of the bow and could not account for the supernumerary arcs, the additional, fainter arcs separated by darker regions that appear inside the main rainbow (Fig. \ref{SupernumeraryPhoto}).
This phenomenon was documented as early as the 13th century \cite{Boyer}. 
%In Figure  we show a picture taken by one of us (A.R) that show these arcs. 

Two important concepts were central to Airy’s treatment: the Huygens concept of wavefronts,  and the mathematical concept of caustic surfaces—the envelope of a family of reflected or refracted light rays\cite{Eckman}. A common example of a caustic curve, a plane section of a caustic surface, are the bright arcs  on the bottom of a swimming pool formed by light rays refracted at the surface.
%\textcolor{blue}{lo dejaria así}.
Applying these concepts to rainbows, Airy derived an analytical expression for the light intensity at each point of the bow and showed that the brightest region lies at a radius smaller than what geometrical optics predicts.
%Airy derived a precise analytic expression for the intensity of illumination at each point of the area brightened by the bow. He found that the intensity of light is given by the square of an integral, which has since become known as ``Airy's rainbow integral''.
%\[
%\int_0^\infty  \cos \left(\frac{\pi}{2}\left(w^3-m w\right)\right) dw,
%\]
%where the parameter \(m\) determines 
%in terms of the angular departure of the ray from the Cartesian ray (See Figure \ref{Cartesian}).
%{\color{blue} no me parece apropiado poner ya la fórmula de Airy y el dibujo en la introducción}
%Before Airy, it had been assumed that the intensity of illumination was greatest at the angle for which the deviation is least, that is, along the Cartesian ray or the caustic curve, of which it is the asymptote. However, Airy's calculations showed that the region of greatest brightness lies appreciably within the radius computed on the basis of geometrical theory. %\textcolor{blue}{Esto no es repetitivo?: He noted that the maximum illumination does not occur at the geometrical caustic, but rather on the luminous side of the geometric position of the rainbow (for the primary bow).}
In this paper, we will show using undergraduate-level concepts that Airy's treatment of caustics can also successfully apply to boat wakes and quantum wave packets, revealing a universal principle that underpins both phenomena.

\section{Caustics from cubic wavefronts. }
\label{CubicFrontSection}
In the context of geometric optics, 
caustics are the envelopes of a family of rays along which light propagates. 
%{\color{red} Sugiero acortar esto asi. Hablamos de envelope of rays y of curves. puede confundie}
The detailed analysis of caustics goes back to the studies of burning mirrors and lenses in the Middle Ages.  Astronomer and musician Francesco Maurolico (1494–1575) noted that, due to spherical aberration, light rays hitting a spherical mirror do not converge at a single focal point, but rather on a surface\cite{MarkSmith}. The caustic
surface is the surface to which the light rays are tangent.
%A familiar example is shown in Figure \ref{CausticCoffee}, where parallel rays (typically coming from the Sun or from a distant lamp) reflect off a circular surface. This is for instance what happens when sunlight hits the inside edge of a coffee cup. The envelope of the reflected rays --a mathematical curve called the ``nefroid"\cite{Rojo}-- is a bright line  whose tip is at a distance $R/2$ \textcolor{blue}{TO DO}from the center $C$ of the circular surface of radius $R$.
%\textcolor{blue}{If I understand correctly, the shape shown in Fig. 2 might be seen on the bottom of the coffee cup. That would be because the rays propagate vertically, in addition to the horizontal components shown in Fig. 2. Providing this extra explanation could help readers make sense of the phenomenon. On the other hand, if it is taking the reader far afield from the main point of the paper, then it might be better to remove this section.
%}
%{\color{red}. Ahi corregi la figura, pero es posible que todo este  parrafo --desde A familiar example ... hasta aqui-- pueda eliminarse.  No me opongo si lo eliminas  }
In this section we will discuss the main properties of caustics with a simple generic example using light rays which is applicable to the three physical realizations presented in the rest of the paper.

%\textcolor{green}{Let us extend briefly the discussion from purely geometrical optics} \textcolor{red}{Acá algo me chocó al leer de corrido. Todavía no presentamos nada... ya ya estamos diciendo "vamos a extender la discusión..." Yo empezaría la oración en el 

Let us consider a two-dimensional curve defining a starting wavefront.
Rays emerge perpendicularly at every point of the wavefront, 
and they converge or diverge depending on the wavefront concavity.
Points of constant optic length along the rays define the time evolution of the wavefront. 
We consider a situation in which the starting wavefront generates a combination of convergent and divergent rays. In other words the starting wavefront  has an inflection point. We will now show how this wavefront gives rise to a caustic line.

\begin{figure}[h]
    \includegraphics[width=15cm]{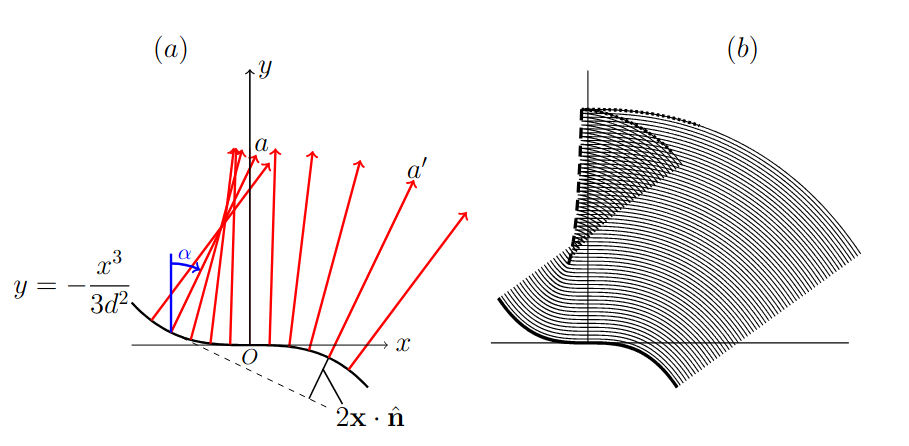}
    \caption{$(a)$: Rays normal to a cubic wavefront parametrized as $\mathbf x=(x,-x^3/(3d^2))$. Rays $a$ and $a'$, originating at $\mathbf x$  and $-\mathbf x$, are parallel and differ in path length by  $2\mathbf{x}\cdot \hat{\mathbf{n}}$, where $\hat{\mathbf {n}}$ is the normal to the wavefront at $\mathbf{x}$. Ray $a$ belongs to a ``fold": it is part of a beam of rays 
    that are initially convergent and then become divergent upon propagation.
    $(b)$: Wavefronts propagating from the initial wavefront shown in panel (a). The structure of these wavefronts, particularly the ``fold" (giving rise to two superimposed divergent fronts (dotted)) and the caustic line (dashed), is  similar to the one observed in the case of raindrops (Fig. \ref{Rainbowfront}b). 
    %The directions of constructive interference (according to Young) are indicated.
    %\textcolor{red}{No sé si esto ponerlo acá. $\to$ As it crosses the caustic it picks %up an extra $\lambda /4$ over the ``bare" path length difference $-2\mathbf{x}\cdot %\hat{\mathbf{n}}$  with ray $a'$.}  }
%\textcolor{blue}{fijate}
    }
    \label{CubicFront0}
\end{figure}

 %Let us consider waves emerging in phase from points along a two-dimensional curve (the wavefront).  
 %These rays propagate perpendicularly to the wavefront and converge or diverge depending on the wavefront concavity.
% The geometric optics description of this situation corresponds to a family of rays that originate at the front and propagate perpendicularly to it. The rays will emerge convergent or divergent depending on whether the front is concave or convex. 
%We consider a situation in which the starting wavefront generates a combination of convergent and divergent rays. In other words the wavefront  has an inflection point and gives rise to a caustic line.
The simplest continuous curve that exhibits this behavior is a cubic wavefront, which happens to arise naturally in the physical phenomena discussed in this paper. It is also the generic form of a smooth curve sufficiently close to an inflection point.
We show an example of such a wavefront in
Fig. \ref{CubicFront0}(a). Note that when the wave has propagated sufficiently far away
from the initial cubic line, the wavefronts ``fold", giving rise to two divergent wavefronts of equal optical paths (Fig. \ref{CubicFront0}(b), dash-dotted lines), that meet at the caustic line (dashed line). 
The folded wavefront also means that, for any direction emerging from the wavefront, two parallel rays exist,
 as illustrated with the rays $a$ and $a'$ in 
Fig. \ref{CubicFront0}(a).
%in this case, where the rays propagate in vacuum, the optical paths are the lengths of the normal line measured from a point of the front to the original front. 
%The optical path $\ell $ of a ray propagating in a medium of variable index of refraction $\mathbf{x}$ is $\ell=\int ds n(\mathbf{x}) $, the length weighted by the index of refraction at each point of the path. 
%\textcolor{red}{Me suena descolgada esta frase $\to$}\textcolor{blue}{And, in geometric optics, the (somewhat paradoxical) term `wave front' is used for the surface of constant path length
%{\bf {OK LA SAQUEMOS}}. Lo puse porque ma parecia una forma de justificar estamos suponiendo que dos rayos cuyo angulo varia cuadraticam,ente como un parametro puede generarse con un frente que depende cubicamente con ese parametro.  
%. }

Near the inflection point, the starting wavefront can be described with the equation $y=-x^3/3d^2$, with $d$ a scale factor (see Figure 3(a)). Equivalently, in parametric form we have: 
\begin{equation}\mathbf{x}(x)=\left(x, -\frac{x^3}{3d^2}\right).
\label{cubicfront}
\end{equation}
The unit normals $\hat{\mathbf{n}}\equiv(n_x,n_y)=(\sin \alpha, \cos \alpha)$ to the curve $\mathbf{x}(x)$ correspond to the directions of the light rays. The angle  $\alpha $ of the normal with the vertical is  given by:
\begin{equation}
\alpha\simeq\tan \alpha = \frac {n_x}{n_y} = \left( {x\over d}
\right)^2.
\label{tg}
\end{equation}

Considering the situation in which rays impact a very distant screen (or  are observed by a distant eye),  the coordinate on the screen is simply  proportional to $\alpha$. 
%{\color{red} As we will see in the next section, for the case of the rainbow , ``distant" means that the observation point is at a distance to the eye much larger than the drop diameter. } 
%\textcolor{blue}{IF this is indeed what you mean, please explicit it in the text : it is not trivial that you consider a screen at infinity since you mentioned earlier the example of a coffee cup where caustics arise precisely because you are close to the reflecting surface.}
Disregarding interference among rays,
the intensity observed for each direction $\alpha$ is given by the density $I(\alpha)$ of rays as a function of $\alpha$. The 
quadratic dependence of $\alpha$ on $x$ (for small $x$) 
%$\alpha\sim u^2$ dependence at small $\alpha$ 
of Eq. \ref{tg} shows no solutions for $\alpha <0$; in other words,  no rays will emerge with $\alpha<0$, meaning that the   intensity is zero  for negative angles.   If we consider the number of rays per unit length of the wavefront $I_0$ as uniform, the number of rays  $dN$ incident in an interval $dx$ is $I_0 \sqrt{1+\alpha^2}dx$.  These rays  emerge in the  interval $(\alpha, \alpha +d\alpha)$: $dN= I(\alpha) d\alpha$, with $I(\alpha)$ the density of rays per unit angle. Using Equation 
(\ref{tg}) we obtain, for small $\alpha$
\begin{equation}I(\alpha)=I_0 \sqrt{1+\alpha²}{dx\over d\alpha}\simeq \frac{I_0d}{2\alpha^{1/2}}.
\label{intensity}
\end{equation}
%Equations (\ref{intensity}) and (\ref{tg})   lead to an intensity given by
%\begin{equation}
%I (\alpha)=\frac{I_0d}{2\alpha^{1/2}},
%\end{equation} 
The above expression shows a divergent accumulation of rays 
 at the angular position of the caustic: 
 $\alpha\to 0^+$.
%\textcolor{blue}{Now, the derivation is more clear to me. However, I would expect the angular density of rays to be the number of rays having the same alpha, no? And, I would have expected the unit to be some intensity/rad. Here you have an additional length. Why is that?} \textcolor{green}{Reply: In fact, $I_0$ has units of number of rays per unit length along the wavefront, so $d$ cancels this length, and $I(\alpha$) is nuber of rays per unit angle (i.e., rad).}
In the next section we will see how the previous generic description applies to the case of the rainbow.
We will then detail the additional improvements that led Airy to quantitatively describe the light intensity in a rainbow.

\section{Airy's (and Young's)  theory of the rainbow. }

\begin{figure}[h]
%   \begin{center}
% \vspace{7.cm}
 %   \begin{tikzpicture}
 %   \node at (1,1) 
    {\includegraphics[width=8cm]{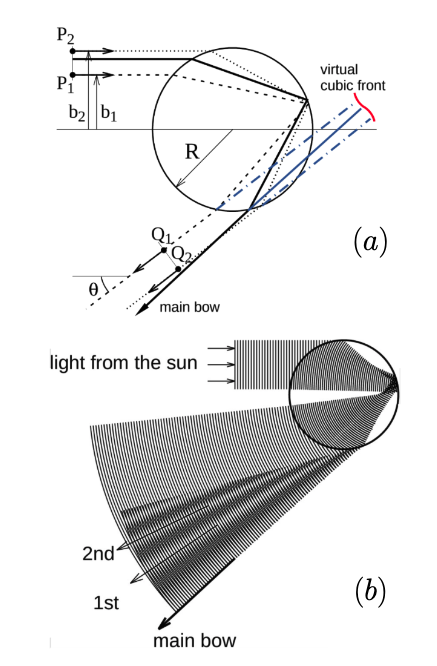}}
%\end{tikzpicture}
%\end{center}
% \hspace{4cm}\includegraphics[width=12cm]{figura_rayos.pdf}\\
%\hspace{1.7cm}\includegraphics[width=12cm]{FiguraRayosB}\\
% \includegraphics[width=6.8cm]{RaysWithCubic.png}
%    \includegraphics[width=8cm]{frente1.pdf}
%  \includegraphics[width=8cm]{frente1A.png}
    \caption{ (a) Rays travelling through a drop, to form a rainbow. The caustic ray of minimum deviation %$\theta_0$ 
    is indicated by the thicker continuous line.
    We also show two additional rays emerging along the same direction, that entered the drop on both sides of the caustic ray (dashed and dotted lines). Since the optical lengths $P_1Q_1$ and $P_2Q_2$ are different these two rays interfere. We show in red the virtual cubic wavefront from which two --in phase-- rays (dash-dotted) originate and give the same path difference. (b) Wavefront description. In rainbows, the sunlight arrives on a droplet with planar wavefronts which are distorted after being submitted to one or several reflections within the droplet. Note the ``folded" nature of the wavefront that comes out of the droplet.    
    The observation directions of the main bow and of the first two supernumeraries (according to Young, see the discussion in text) are indicated.  
    \label{Rainbowfront}
}    
\end{figure}

In the geometric optics description, the primary rainbow forms when sunlight undergoes two refractions and one reflection inside a spherical water droplet (Fig. \ref{Rainbowfront}). 
The key point in the formation of the rainbow  is that the angle \(\theta\) between the incident ray and the ray emerging from the droplet varies non-monotonically with the distance \(b\) of the incident ray to the droplet’s center.  This non-monotonic dependence gives rise to a caustic.

Specifically, if we define the dimensionless length \( z \equiv \frac{b}{R} \)  where \( R \) is the radius of the droplet, a straightforward geometric calculation using Snell's law yields

\begin{equation}
\theta(z)=4 \sin ^{-1}(z/ n)-2\sin ^{-1} z, 
\label{ThetaOfx}
\end{equation}
where $n$ is the refractive index of the droplet.  This function has a minimum for
\begin{equation}
    z_0=\sqrt{4-n^{2}\over 3},
    \label{rainbowangle}
\end{equation}
which, for  $n\simeq 1.33$, gives the well-known angle of the principal rainbow $
\theta_0 \simeq 42 ^{\circ}$. 
In other words, the rainbow appears at an angle corresponding to a sharp boundary between  a dark and a bright region where light rays concentrate. This angle therefore corresponds to a caustic, as described in the previous section.
Differences in the value of $n$ for the different wavelengths are responsible for the spread of colors in the  rainbow.

If we stay within the geometric optics description, the intensity decreases monotonically as the viewing angle deviates from $\theta_0$.
To compute the ray intensity as in Equation (\ref{intensity}), where we now have $\alpha=\theta-\theta_0$,
we expand the angle between rays around the extremum
\begin{equation}
\alpha \simeq  \frac{1}{2}\theta''(z_0)\delta^2,
\label{quadratic}
\end{equation}
with $\delta\equiv (z-z_0)$ and
\begin{equation}
\theta''(z_0)=-{9\over 2} {\sqrt{4 - n^2}\over({ n^2-1})^{3/2}},
\end{equation}
and obtain for the ray intensity $I$  close to the caustic:
\begin{equation}
I(\alpha)= \frac {I_0}{\left |d\alpha/dz\right |}=\frac{I_0}{|\theta''(z_0)|\delta}= \frac{I_0}{\sqrt{2\theta''(z_0)\alpha}}
\label{Intensity}.
\end{equation}
In the present case we have a family of initially parallel rays (perpendicular to the vertical line $P_1P_2$ of Figure \ref{Rainbowfront}) that  emerges from the droplet forming a small angle $\alpha \sim {1\over 2}\theta''(z_0) \delta^2$ with the direction of the caustic ray. 
For each value of the exit angle $\alpha$, there are two light rays contributing, as discussed in the previous section.
%We have two rays emerging with the same angle, as in the discussion in the previous section. 
In the latter case, the dependence of $\alpha$ on $x$ (Eq. (\ref{tg})) originated in the form of the cubic wavefront:
$y=-x^3/(3d^2)$. 
Note that, since  Eqs. (\ref{cubicfront})  and (\ref{tg}) can be written in
 terms of a dimensionless variable  $\tilde \delta\equiv x/d$  (as  $y=-d\tilde\delta^3/3$ and  $\alpha=\tilde\delta^2$), the dependence of $\alpha$ on $\delta$ for the rainbow (Eq. (\ref{quadratic})) can be thought as originating from a virtual cubic wavefront (Fig. \ref{Rainbowfront}a)
given by
\begin{equation}
y=- {R\over 6}  \theta''(z_0)\delta ^3,
\label{CubicVirtual}
\end{equation}
where $y$ is measured along the direction of the rainbow´s main bow (See Figure \ref{Rainbowfront}a).

The existence of two rays for each observation direction, combined with the assumption of the wave-like nature of light, provided the key ingredients to address the phenomenon of ``supernumerary arcs."  This problem was first tackled in the early 19th century by Thomas Young and later refined by George Airy, as we will discuss in the following sections.

\subsection{Young's theory }
Thomas Young provided the first 
explanation linking the supernumerary arcs to interference effects\cite{Young}. 
Young attributed the dark spaces between supernumerary arcs to destructive interference between the two rays emerging from each water droplet in the same direction. 
Two such rays are indicated in Fig. \ref{CubicFront0} as $a$ and $a'$. Their difference in optical path $\ell$ is given as
\begin{eqnarray}
    \delta\ell=|2\mathbf{x}(x)\cdot \hat{\mathbf{n}}|&=&
    \left|2 \left(x\sin \alpha-{x^3\over 3d^2}  \cos\alpha 
    \right)\right|
        \\
    &\simeq&{4\over 3}d \alpha^{3/2},
    \label{Young}
\end{eqnarray}
where where $\hat{\mathbf{n}}$ is the normal to the wavefront at $\mathbf{x}$, and where we used $\sin \alpha \sim \alpha$ and $\cos \alpha \sim 1$.
In the case of the virtual cubic front emerging from the droplet:
%Applying this analysis to our actual cubic front (Eq. \ref{CubicVirtual}) we obtain for the drop
\begin{equation}
\delta \ell
%=\ell(x_1)-\ell(x_2)
= {4\over 3} R \sqrt{2\over \theta''(z_0)} \alpha ^{3/2}.
\label{ll}
\end{equation}
Constructive interference arises when the path difference satisfies 
$
\delta \ell = m\lambda,
$
with \(\lambda\) denoting the wavelength of light and \(m\) an integer. 
The intensity is then modulated by a factor of 
$
\cos^2\left({\pi \delta \ell}/{\lambda}\right).
$
 The
Young's result for the intensity is obtained as (using Eqs. \ref{ll} and \ref{Intensity})

\begin{equation}
I(\alpha)=  \frac{I_0}{\sqrt{2\theta''(z_0)\alpha}}\cos^2\left (\frac {4\pi R}  {3\lambda}  \sqrt{2\over \theta''(z_0)}\alpha^{3/2}\right ).
\label{Young_completa2}    
\end{equation}
We plot this intensity in Figure \ref{AiryOriginal}.
Notice the existence of supernumerary arcs, separated by points of zero intensity, corresponding to the destructive 
interference condition. The main arc appears as $\alpha\to 0^+$, where the intensity diverges. This unrealistic 
behavior is superseded by Airy's theory that we discuss in the next section. 

\subsection{Airy's theory of supernumerary arcs}

%Consider the two rays emerging at a small angle $\alpha$ with respect to the caustic ray. To lowest order these were originated in two rays entering the drop at $x_1=x_0-...\alpha^{1/2}$ and $x_2=x_0+...\alpha^{1/2}$.

%Incident rays that cross the vertical line $AB$ in Figure \ref{Rainbowfront} are in phase for different values of $x$, and when they emerge from the drop and cross the line $CD$ they have a phase difference $\Delta \phi(\delta) =2k R\theta''(x_0) \delta^3 /3$ with $k=2\pi / \lambda$. 

%The description we are obtaining for the rainbow problem is in fact characteristic and generic of diffraction effects near caustics.
%A simpler geometrical argument can be given, realizing

Whereas Young treated the problem of the supernumerary arcs as an interference effect between two parallel rays that propagate perpendicular to the source wave front, Airy considered that every point on the wavefront generates rays in all possible directions (i.e., for all values of $\alpha$) and computed the amplitude $U$ as the sum of the contributions from all source points.

Following Airy, we consider rays with a given $\alpha$ that emerge from every point of the cubic front (Fig. \ref{CubicFront0}). All of them are focused at the same point when observed from a very large distance $\boldsymbol{\rho}=\rho \hat{\boldsymbol{n}}$, and we should account for the interference among all of them.
The optical path of these rays is given by
\begin{equation}
    \ell=|\boldsymbol{\rho}-\mathbf{x}| \simeq \rho -\hat{\boldsymbol{n}}\cdot \mathbf{x}\simeq \rho-\alpha x+ {1\over 3} {x^3\over d^2},
\end{equation}
where $\rho$ is the reference optical length of the ray corresponding to $x=0$ once it has propagated to the observation point.
 The amplitude $U(\alpha)$  is therefore given by the superposition of plane waves that originate from different points on the wavefront. This is essentially the content of Huygens's principle.
 We obtain 
    \begin{eqnarray}
   U(\alpha)&=& e^{ik \rho}\int_{-\infty}^{\infty} dx e^{-ik \left( \alpha x - {1\over 3} {x^3\over d^2}\right)}  \label{AiryCentral}\\
  &=& 2\pi\left(d^2\over k\right)^{1/3} e^{ik \rho} {\rm{Ai}}\left(-(kd)^{2/3}\alpha\right )
   \label{AiryCentral2}
 \end{eqnarray}
 where $k\equiv 2\pi/\lambda$,   $d=R\sqrt{2/|\theta''(z_0)|}$ and ${\rm{Ai}}(x) \equiv \frac 1 \pi\int_0^{\infty}du\cos(u x+u^3/3)$
 the celebrated Airy function.  
 \begin{figure}[h]
    \includegraphics[width=8cm]{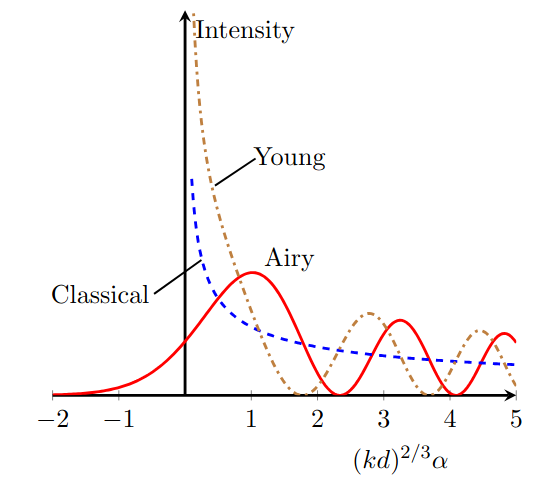}
    \caption{ Rainbow intensities as a function of the angle $\alpha\equiv \theta-\theta_0$ from the ``classical" angle $\theta_0\simeq 42^{\circ}$ , for the different approaches  presented in the text. The dashed line is the result  of Eq. (\ref{Intensity}), which completely disregards the wave nature of light.  Young's curve corresponds to  Eq. (\ref{Young_completa2}), and Airy's result  is the one of Eq. (\ref{AiryCentral2}). Notice that away from $\alpha=0$, Young's  and Airy's approaches give a remarkably similar form, although the maxima are shifted. This shift is related to the so-called Gouy's phase\cite{Gouy}, which is a phase shift that occurs when rays go through a caustic or a focus. See the Supplementary Material for a brief account of this effect.
    }
    \label{AiryOriginal}
\end{figure}
{
To better understand the form of the Airy function \cite{stat_phase} {(see also Appendix I)}, consider the case where \(\alpha > 0\). In this scenario, the argument of the exponential (denoted as \(\Theta\)) in Eq. (\ref{AiryCentral}) has two critical points, \(u_\pm = \pm d\sqrt{\alpha}\), where \(d\Theta/du = 0\). Since \(\Theta\) is stationary at \(u_\pm\), the regions around these points provide the dominant contributions to the Airy function.  
Approximating the integral by focusing on these contributions is known as the method of stationary phase, or saddle point approximation (see Appendix I and Ref.\cite{stat_phase}).
The contributions from the regions around \(u_+\) and \(u_-\) differ in phase, with a phase difference depending on the value of \(\alpha\). As a result, the Airy function oscillates when \(\alpha > 0\).

On the other hand, for \(\alpha < 0\), there are no points of stationary phase, and \(\Theta\) varies significantly with \(u\) throughout, leading the Airy function to be exponentially attenuated. The Airy integral also provides a smooth transition between the oscillatory regime for \(\alpha > 0\) and the exponentially decaying regime for \(\alpha < 0\).
}   

In Figure \ref{AiryOriginal} we show the intensity according to Airy's (Eq. (\ref{AiryCentral2})) (the intensity is obtained as $I=|U|^2$) and Young's (Eq. (\ref{Young_completa2})) results. One important success of Airy's treatment is to provide a finite intensity at the caustic edge.
In addition, it shows that the maximum intensity does not coincide with the predictions of geometric optics. It also accurately
accounts for the angular position of the supernumerary arcs that depends on the radius of the droplet, and are incorrectly predicted in Young's treatment (see the Supplementary Material for a discussion of this difference).
The droplet radius $R$ affects directly some observable characteristics of the rainbow.
Airy's result shows that each wavelength produces a disk of light that is brightest at its caustic edge and fades inward. 
The angular width $\Delta \alpha$ of the disk is approximately (Eq. (\ref{Young_completa2}) and Fig. \ref{AiryOriginal}) $\Delta \alpha \simeq (\lambda/R)^{2/3}$.
These disks overlap substantially for different wavelengths, all the more as $R$ is small, mixing colors and reducing spectral purity. As $R$  increases, the arcs of different colors become more sharply defined and less overlapping, resulting in a brighter and more vividly colored rainbow.
This is why the most vivid rainbows typically appear during the warmer months, especially after an afternoon thunderstorm, 
 as such conditions favor the formation of large raindrops. 
In fact, thunderstorms involve intense upward air currents (convection), which suspend droplets in the atmosphere longer, allowing them to grow larger by merging with smaller droplets. 
The dependence on the raindrop radius also explains the fact that
supernumerary arcs are more prominent  at the bow's apex\cite{Boyer2}. In fact, since raindrops increase in size while falling, they are the smallest at the central point of the rainbow, and this produces the largest angular separation between supernumerary arcs, according to Airy's formula.

Airy's treatment is widely regarded as the definitive theory of the rainbow, primarily because it explains most of observed properties under realistic conditions. However, it is important to note that Airy's theory still describes light propagation in terms of rays. This approach has its limitations, especially when the size of raindrops becomes comparable to the wavelength of the incident light. In such cases, a more complete wave theory—specifically, one based on the Maxwell equations of the electromagnetic field—is needed. For a spherical drop, this is captured by Mie's scattering theory\cite{Adam}, which, among other things, predicts that rainbows tend to appear fuzzier and more whitish when the raindrops are very small.

\section{Airy oscillations in boat wakes}

%\textcolor{blue}{ver comentario de Beth}
The treatment of the rainbow presented in the previous section is found in a variety of texts. 
In contrast, the analysis of boat wakes {(originally considered by Kelvin\cite{Kelvin1})}
is less common and has even recently sparked controversies\cite{Froude}.
The application of Airy's theory to the Kelvin wake problem is also rarely discussed, although it can be found in Ref. [15]. %\cite{Lighthill}. 
We present the Kelvin wake problem 
in a way that highlights its analogy with the formation of rainbows.

Consider a nearly point-like source that creates a disturbance on the surface ($Oxy$) of a deep liquid. The source moves with uniform velocity $v$ along a direction that we define to be the $x$ axis. We want to determine the most general expression for the perturbation in surface height $h(x,y)$ that this source can produce. 
To solve the problem, we consider the two half-planes $y>0$ and $y<0$ separately, using the $y=0$ line along which the boat moves as a boundary condition.
In each half-plane, we
decompose the surface perturbations into plane waves, taking advantage of
the fact that, in the frame moving with the source, the resulting pattern should be stationary. 

Each plane wave of  wave vector ${\bf k}\equiv(k_x,k_y)$ corresponds to a perturbation that is proportional to (the real part of) $\exp[i(k_xx+k_yy-\omega({\bf k}) t)]$, with $\omega$ given by the medium's dispersion relation.  For deep-water waves it reads\cite{Crawford} 
\begin{equation}
    \omega({\bf k})=\sqrt{g|{\bf k}|},
    \label{dispersion}
\end{equation}
with $g$ the gravitational acceleration. 
In order for the amplitude to be stationary in the frame of reference moving with the source, the phase 
of the plane must be proportional to $(x-vt)$.  In other words, since the height $h$ of the disturbance must satisfy $h(x,y,t)= h(x-vt,y,0)$, we have the following relation:
\begin{equation}
\exp\left [ i(k_xx+k_y y -\omega(\mathbf{k})t\right]=
\exp\left [ i(k_x(x-vt)+k_y y \right],
\label{integral_kelvin0}
\end{equation}
which is satisfied if 
\begin{equation}
k_xv=\omega=g^{1/2}(k_x^2+k_y^2)^{1/4}.   
\label{Univocally}
\end{equation}
Then, $k_y$ can be expressed in terms of $k_x$:
\begin{equation}
k_y=\pm k_x\sqrt{\frac{k_x^2v^4}{g^2}-1}.
\end{equation}
We remark that, from Eq. (\ref{Univocally}),  $k_x$ and $v$ must have the same sign and we take both as positive.
Also, when considering the $y>0$ ($y<0$) half plane, we must have $k_y>0$ ($k_y<0$) in order to have 
plane waves that move away from the boat.
From now on, we focus on the $y>0$ half-plane, measure $x$ and $y$ in units of $v^2/g$, and use $k$ to refer to $k_x v^2/g$. 
The most general form of a surface disturbance then is:
\begin{equation}
h=\int_{1}^{\infty} dk A_k e^{ i\left(kx+k\sqrt{k^2-1}y\right)}.
\label{integral_kelvin}
\end{equation}
The lower limit of the integral is 1, because lower values of $k$ produce waves that exponentially attenuate for $|y|\to\infty$, and we disregard them. The coefficients $A_k$ and the exact value of the upper integration limit are important in determining the actual details of the wake for a boat of given shape and dimensions. 
As we are interested in a point source we can take $A_k\equiv 1$, and the upper limit as infinite. 

There is a strong formal similarity between Eq. (\ref{integral_kelvin}) and Eq. (\ref{AiryCentral}) that reveal a striking analogy between the supernumerary arcs and the water wake pattern.
%At first glance, the connection with the rainbow may not seem obvious. Our description of the water wake focuses on its wave nature, rather than using a ray-based approach. Additionally, the wake amplitude results from the superposition of waves with different wavelengths, whereas the rainbow pattern arises from a monochromatic wave. 
To uncover this similarity, we first evaluate Eq. (\ref{integral_kelvin}) with the saddle point method.
For large values of $x$ and $y$, the argument of the integral in  Eq. (\ref{integral_kelvin}) is rapidly varying, and Eq. (\ref{integral_kelvin}) gets its main contribution from $k$-values where the phase is stationary: $\partial ({kx+k\sqrt{k^2-1}}y)/\partial k=0$. These points are

%\textcolor{blue}{As in the rainbow case, Eq. (\ref{integral_kelvin}) gets its main contribution from points in which the phase varies the least with the integration variable. To identify these points 
%we look for the values of $k$ for which $\partial ({kx+k\sqrt{k^2-1}}y)/\partial k=0$, and obtain}

\begin{equation}
     k_\pm (x,y)= -{1\over \sqrt{8} }{x\over  y}  \sqrt{4+\left({y\over x}\right)^2 \pm \sqrt{1-8\left({y\over x}\right)^2}}.
     \label{ValuesOfK}
\end{equation}
%\begin{equation}
%\frac xy=\frac{1-2\bar k^2}{\sqrt{\bar k^2-1}},   
%\label{fase_est}
%\end{equation}
%\textcolor{blue}{las dos soluciones son por el sino de la raíz interior, no la de afuera... ///// Sí, son cuatro soluciones en realidad. Dos para cada lado del eje y.  )}
Equation (\ref{ValuesOfK}) provides either two or no values, depending on whether \( x/y \) is less  or greater than $-\sqrt{8}$. This condition defines an angle $\theta_K=\tan^{-1}(1/\sqrt{8})=19.5^\circ$, which is the angle of the famous Kelvin's cone observed as a boat moves. Note that the value of $\theta_K$ is totally independent of  the boat's velocity.
The cone appears as a caustic, separating a region with two phase-stationary solutions from another ``dark" region without any possible stationary phase solution.

\begin{center}
\begin{figure}
%\begin{tikzpicture}
%\includegraphics[width=15cm,clip=true]{kelvin_pattern.pdf}
\includegraphics[width=8cm,clip=true]{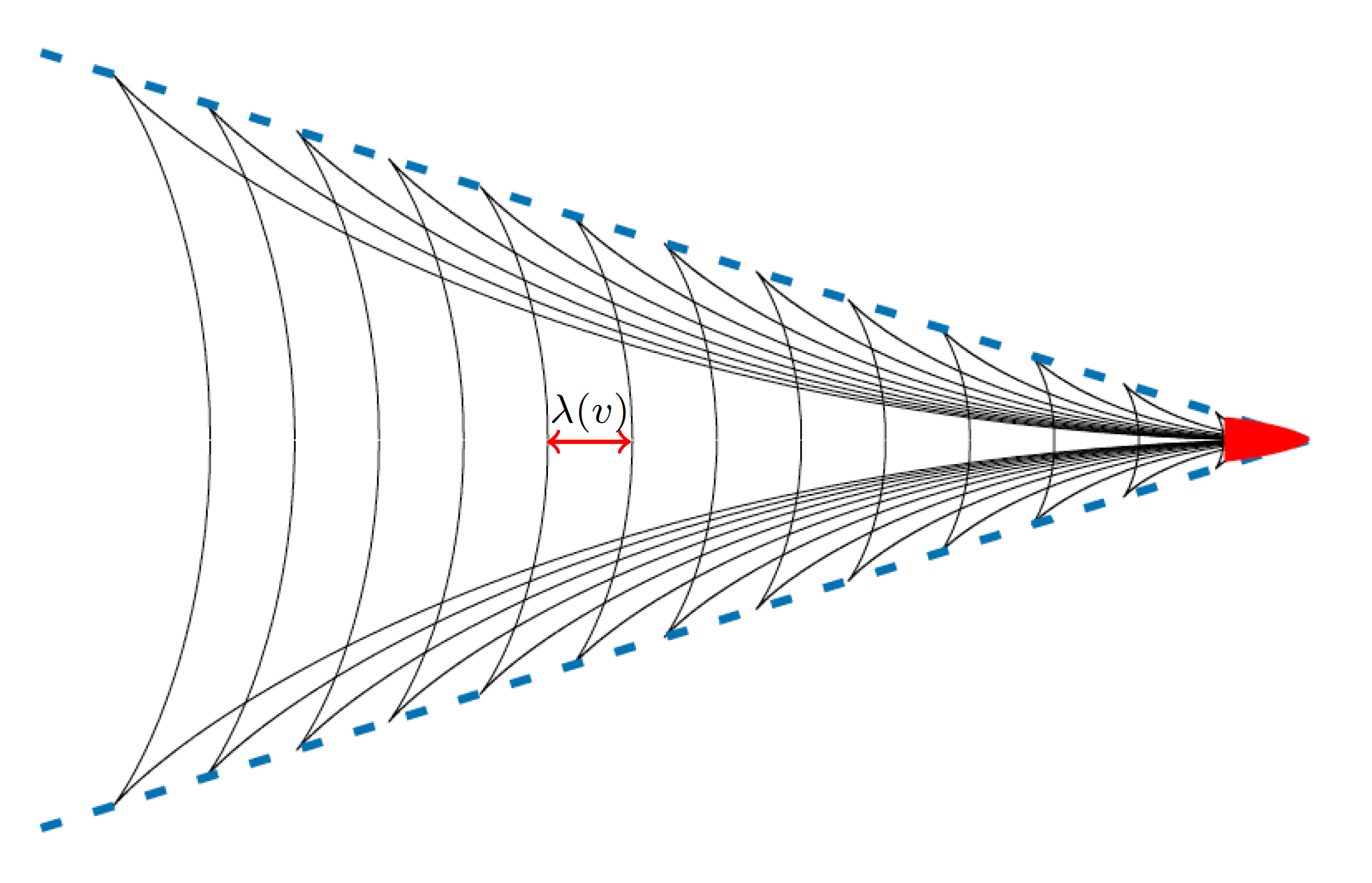}
%\node at (0.,2) {$\lambda$};
%\end{tikzpicture}
\caption{For a boat at the right tip, moving towards the right, we show with continuous lines lines the positions of the maxima of perturbations (crests) as obtained within the stationary phase approximation. There are no solutions outside the Kelvin cone (dashed line), with half angle 19.5$^\circ$ (dashed lines). 
{
This diagram contains some of the essential features exhibited by the observed wake. The angle of the cone (the caustic) matches that observed in real boat wakes. In addition, inside the cone there are arcs corresponding to a slightly curved plane wave whose phase velocity matches the boat velocity $v$. This is clearly observed in Fig.~\ref{cambio_de_fase}. Using Eq.~(\ref{dispersion}), we see that the velocity dependent  wavelength $\lambda(v)$ of this portion of the pattern is $\lambda(v)=  2\pi v^2/g$. Once diffraction effects near the caustic are included, we obtain a more realistic pattern near the edge of the wake that includes a finite wave amplitude beyond the cone, in close analogy to what is observed in the case of the rainbow (See Fig. \ref{kelvin_completo}).
}}
\label{f2}
\end{figure}
\end{center}

Re-inserting the values of $k$ from Eq. (\ref{ValuesOfK}) back into the phase of Eq. (\ref{integral_kelvin}), we obtain an  oscillatory  envelope $\exp\left\{i \left(k_{\pm}x +k_{\pm}\sqrt{k_{\pm}^2-1}y\right)\right\}$  of $h(x,y)$ in the stationary phase approximation. 
The maxima of these perturbations are plotted in Fig. \ref{f2}.
%This is the result first obtained by Kelvin. In this approximation the Kelvin cone is a sharp boundary between regions of zero and finite wave amplitude.
%However,  this is only an approximate result that is modified when the integral in Eq. (\ref{integral_kelvin})
%is exactly evaluated. \textcolor{blue}{No me gusta como suena esta frase
%}
This result, first obtained by Kelvin, predicts that the Kelvin cone acts as a sharp boundary separating regions of zero and finite wave amplitude. However, this sharp boundary is only an artifact of the approximation; an exact evaluation of the integral in Eq. (\ref{integral_kelvin})  reveals a smooth crossover between the two regions.

%Plotting the lines where this phase if a multiple of $2\pi$, the results in Fig. \ref{f2} is obtained, 
%which is for the Kelvin problem.

\begin{figure}[h]
\includegraphics[width=15cm,clip=true]{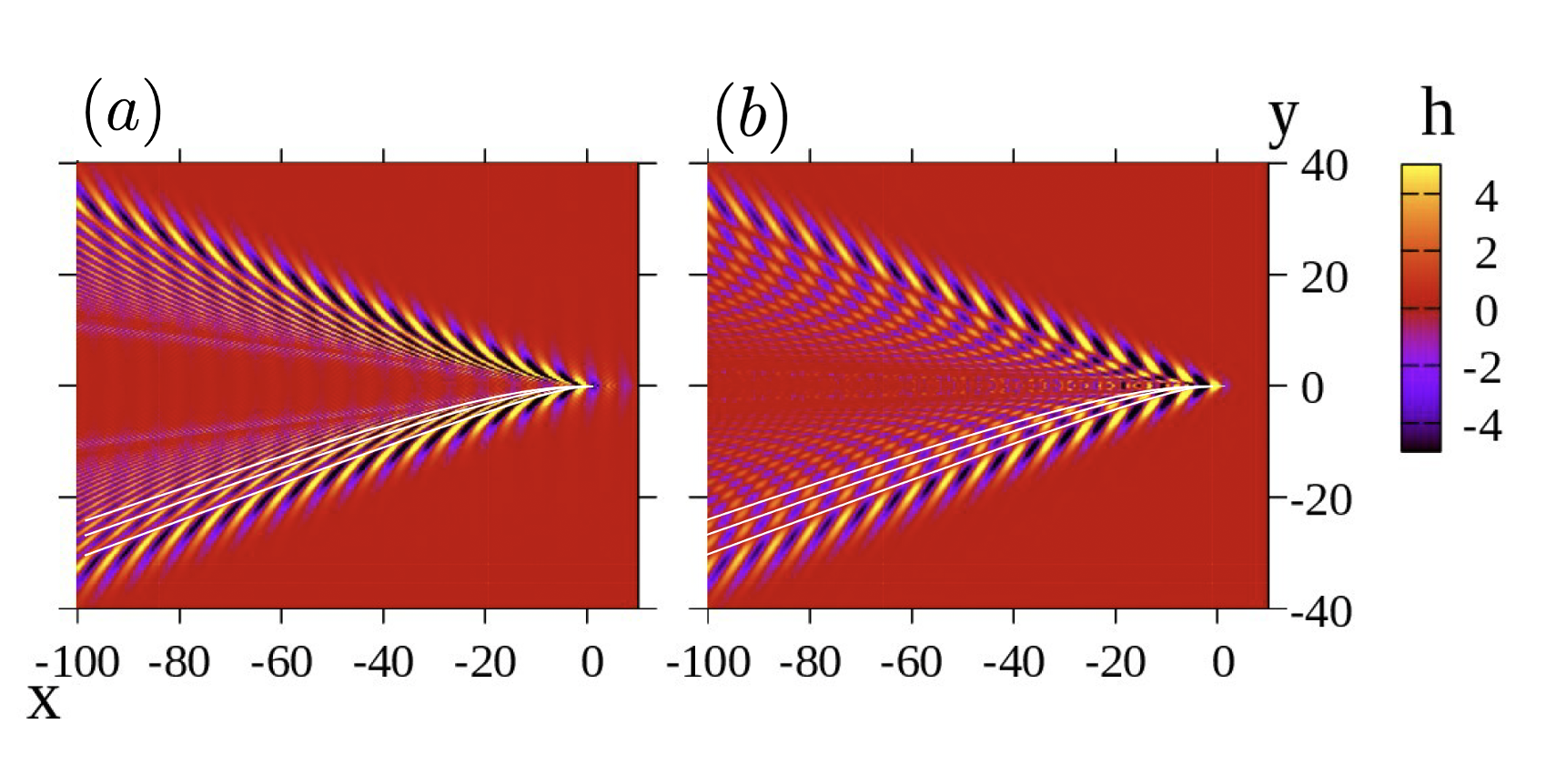}
\caption{(Color online) Surface perturbation $h$ produced by point source at $(0,0)$ moving to the right. (a) is the exact result of Eq. (\ref{integral_kelvin}), obtained through numerical integration ($A_k\equiv1$). (b) is the analytical result of Eq.  (\ref{airycos}) ($C\equiv 1$), which is expected to provide accurate results near the Kelvin cone, and away from the source. In the lower part of both panels we plot as white lines the positions of zero amplitude from Eq. (\ref{zeros}), corresponding to the first three zeros of the Airy function.}
%\textcolor{blue}{
%Is there a reason that the right panel shows maxima of the perturbation that are end in a triangle (as I've tried to represent on the schematics) and not as arcs like in the left panel and what you drew in Fig. 6? I understand the approximation is less exact away from the cone, but if Fig. 6 shows them as circles, we should recover that in Fig 7, no?}\textcolor{red}{Response. The only reason is that the right panel is accurate only near the Kelvin cone. The crests showed in Fig 6 are obtained from the stationary phase condition, and are accurate for all the interior of the cones. That is why it compares better with the left panel.}
%}
\label{kelvin_completo}
\end{figure}

\begin{figure}[h]
\includegraphics[width=8cm,clip=true]{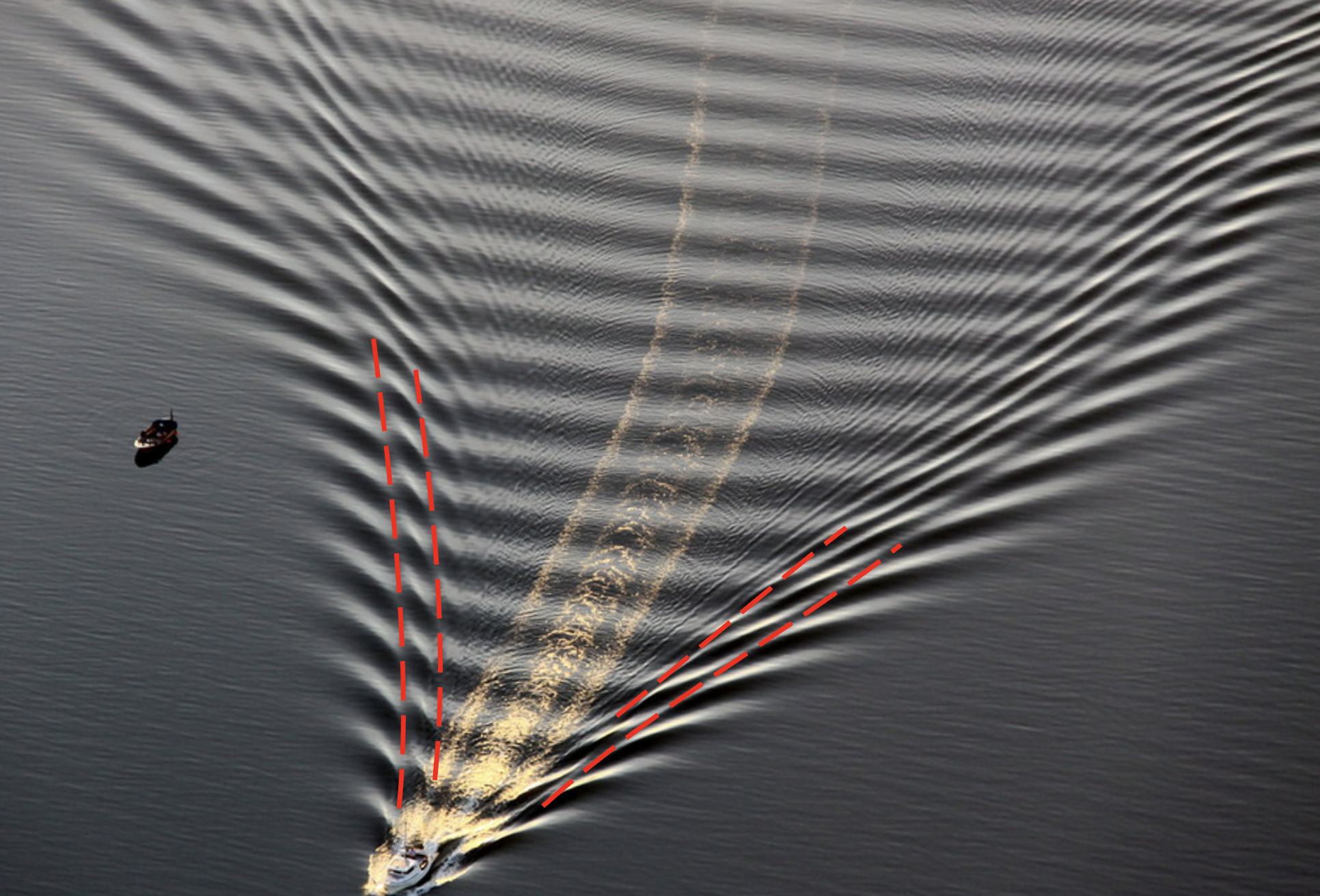}
\caption{(Color online) Real life example of a ship wake in which a few lines of almost zero perturbation (corresponding to the zeros of the Airy 
function in the asymptotic expression of Eq. (\ref{airycos})) are highlighted.
\textcolor{black}{(Photograph by Chris Goldberg, reproduced with permission.)}
}
\label{cambio_de_fase}
\end{figure}

%In addition, just as in the case of the supernumerary arcs,  there is a smooth transition between the two regions, that can be obtained through an analysis similar to that of Airy for the rainbow.
%\textcolor{blue}{Could you comment a bit more on fig 6?
%* Kelvin's cone and comparison with experiment
%* Distance between these maxima and comparison with experiment. Does one sees circular waves (are they truely circular) within the %cone? What is the center (focal point) of these waves? Is this focal point analogous to that of rainbows?
%* What about the waves that fold close to the cone?}
To evaluate Eq. (\ref{integral_kelvin}) analytically near the Kelvin cone,
one should note that the phase in the integral in Equation (\ref{integral_kelvin}) has an inflexion point at $k_0=\sqrt{3/2}$,
at which its second derivative vanishes. 
Expanding around this point, we obtain  
\begin{equation}
    kx+\left(k\sqrt{k^2-1}\right)y\simeq \left(  \sqrt{\frac 32} x+\frac{\sqrt 3}{2}y     \right) + u (x+\sqrt{8}y) + \sqrt{8}y u^3,
\end{equation} 
where $u=k-k_0$. In this way, near $u\sim 0$
\begin{equation}
h=C e^{i\left(  \sqrt{\frac 32} x+\frac{\sqrt 3}{2}y     \right)} \int du     e^{ i\left( 
u (x+\sqrt{8}y) + \sqrt{8}y u^3  \right)},
\label{h}
\end{equation}
with $C$ an overall constant.  
%\textcolor{blue}{Acá Claire dice algo que está mal} 
After integrating over $u$
%
%and considering the deviation $\alpha\equiv \tan^{-1} (y/x)-\theta_K$ from the direction of the Kelvin cone to be small,  
we obtain (taking the real part of the final expression)
\begin{equation}    
h(x,y)\simeq 2\pi C y^{-1/3}\cos \left(  \sqrt{\frac 32} x+\frac{\sqrt 3}{2}y  \right)
\times  {\rm{Ai}} \left(\frac{x+\sqrt 8 y}{(3{\sqrt{8}}y)^{1/3}}\right).
\label{airycos}
\end{equation}  
This expression is the equivalent to Eq. \ref{AiryCentral2} for the rainbow.
Since the pattern is moving in the $x$ direction with velocity $v$, the above equation shows that an observer at rest, at a distance $y$ from the path of the moving boat 
%(the $x$ axis in our description) 
will observe an oscillation of  the water level in time that can be described by an Airy function superimposed with an harmonic factor due to the cosine function in Eq. (25). The overall amplitude is proportional to $y^{-1/3}$.
%\begin{equatin}    
%h\simeq {y^{-1/3}}\cos \left(  \sqrt{\frac 32} x+\frac{\sqrt 3}{2}y \right)
%\times  {\rm{Ai}} \left( {3\over \sqrt{2}}(x^2+y^2)^{4/3} \alpha \right)
%\label{airycos}
%\end{equation}
%where we see that the dependence of $h$ on $\alpha$ is similar to that obtained in Eq. (\ref{AiryCentral2}) for the rainbow.
This result, valid near the Kelvin cone, is plotted in Fig. \ref{kelvin_completo}, together with the exact result obtained by numerical integration of equation (\ref{integral_kelvin}).
%The ``caustic" nature of the pattern at the Kelvin cone is observed in the slower decay of the amplitude $h$ with distance, $h\sim \rho^{-1/3}$, instead of the $h\sim \rho^{-1/2}$ dependence observed well in the interior of the cone.
Note how the Airy function produces fringes of maximum amplitude, which are analogous to the supernumerary arcs in the rainbow problem. 
In between these maxima,
there are lines where the perturbation amplitude vanishes, which correspond to the zeros of the Airy function. 
Given a zero $z_0$ of the Airy function (${\rm{Ai}}(z_0)=0$), the perturbation vanishes along the curve
\begin{equation}
x+\sqrt{8}y=z_0 (3\sqrt 8y)^{1/3}.
\label{zeros}
\end{equation}
These lines of ``destructive interference" %can be guessed in Fig. \ref{f2} (they are indicated by the black lines), as the 
%lines along which the two waves inside the Kelvin cone are in counter-phase, and therefore annihilate. The 
%existence of these lines 
can sometimes be seen and are one of the most delicate and beautiful details of boat wakes (Fig. \ref{cambio_de_fase}).

%They occur  to the left of the Kelvin cone, displaced by an amount $\delta x= C_i y^{1/3}$, where $c_i$ are numerical coefficients associated to the zeroes of the Airy function.
%{\color{red} Esto que sigue es igonarndo Gouy cierto?}{\color{blue} No estoy seguro. Está hecho con fase estacionaria, y eso (en el caso del arcoiris) te da la posición correcta de los máximos y mínimos}

\section{(Quasi) non-dispersive quantum wave packets}

The mathematical structure of Eq. (\ref{integral_kelvin}) suggests its application to yet another physical problem. 
Consider a  quantum particle moving in one dimension in the presence of a periodic potential $V(x)$ generated by identical atoms placed at positions $x=na$, with $n$ being an integer. The eigenfunctions of this problem are characterized by a crystal momentum $p$ with $-\pi\hbar/a<p<\pi\hbar/a$, 
and a dispersion relation  $\varepsilon(p)$. The wave function amplitude at the different atomic positions can be written as a superposition of solutions with different crystal momenta in the form
\begin{equation}
\Psi(an,t)=\int_{-\pi\hbar/a}^{\pi\hbar/a} dp A_p e^{ {i\over \hbar}\left[anp-\varepsilon(p)t\right] }.    
\end{equation}
If a wave packet constructed at $t=0$ has its values concentrated around a well-defined spatial position, then the velocity of this packet is given by the group velocity $v_g=\partial \varepsilon(p)/\partial p$.

%Consider a  quantum particle moving in one dimension, with a dispersion relation (energy $\varepsilon$ versus momentum $p$) given by
%$\varepsilon(p)$.  For a free non-relativistic particle   $\varepsilon(p)=\displaystyle{p^2\over 2m}{}$.

%The time-dependent one-dimensional Schr\"odinger equation for the wave function $\Psi(x,t)$ of the particle is 
%\begin{equation}
% i\hbar \frac {\partial \Psi}{\partial t}=\hat H \Psi,    
%\label{schro0}
%\end{equation}
%with  $\hat H\equiv \hat p^2/2m)$, and $\hat p\equiv -i\hbar\partial/\partial x$.
%The general solution of (\ref{schro0}) can be  written as
%\begin{equation}
%\Psi(x,t)=\int dp A_p e^{ {i\over \hbar}\left[px-\varepsilon(p)t\right] }.
%\label{schro}
%\end{equation}
%where $\varepsilon(p)\equiv p^2/2m$ is the dispersion relation.
%An elementary analysis then shows that the phase velocity $v_{ph}$ of the wave packet of (Eq. (\ref{schro})) is given by $v_{ph}=\varepsilon(p)/ p$, while 
%the group velocity (the velocity of the envelope of the packet) is given 
%by $v_g=\partial \varepsilon(p)/\partial p=2v_{ph}$. 

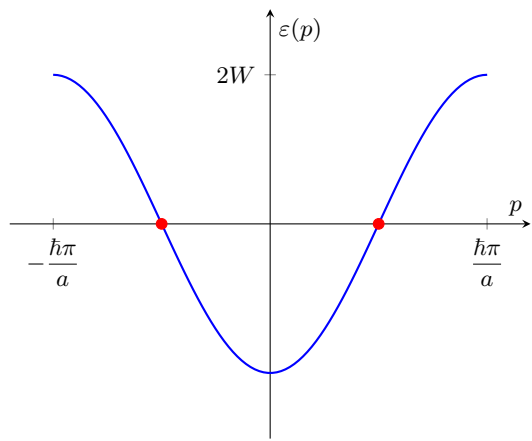
\begin{figure}[h]
\begin{tikzpicture}
    \begin{axis}[
        axis lines = middle,
        xlabel = $p$,
        ylabel = {$\varepsilon (p)$},
        samples = 100,
        domain = -pi:pi,
        xtick = {-6.28, -3.14, 0, 3.14, 6.28},
        xticklabels = {$-2\pi$, $-\displaystyle{\hbar\pi\over a}$, $0$, $\displaystyle{\hbar\pi\over a}$, $2\pi$},
        ytick = { 1},
        yticklabels = {$2W$},
        ymin = -1.2,
        ymax = 1.2,
        enlargelimits=true
    ]
    \addplot[blue, thick] { -cos(deg(x)) };
   \addplot[mark=*, red, only marks] coordinates {(pi/2, 0)};
     \addplot[mark=*, only marks,red] coordinates {(-pi/2, 0)};
    \end{axis}
\end{tikzpicture}
\caption{The energy-momentum relation $\varepsilon(p)$ of the tight binding model of particles in a one-dimensional discrete lattice  has two inflection points, indicated by the small circles. 
%\textcolor{blue}{Please bear in mind that the printed version of the journal will have black and white figures, while the online version can display colored figures (of course). This means that it would be best to edit your figures so that both versions are equally readable.}
}
\label{inflection}
\end{figure}

The spreading of the wave packet is controlled by the  second derivative $\partial^2 \varepsilon(p)/\partial p^2$. 
%A geometrical interpretation of this fact is contained in Fig. \ref{packet}.
In fact, by constructing a wave packet with different $p$ values, slightly different group velocities are incorporated, and therefore
different parts of the packet move at different velocities, producing an overall spreading.\footnote{This argument about the 
spreading of a packet has one assumption, that is rarely made explicit. The components of the packet
at the initial time must be chosen in-phase. If this is not accomplished, one can construct a wave packet that shrinks as a function of 
time down to its minimum size allowed by the uncertainty principle, before it starts spreading again at later times.}
It is natural to ask what happens if a wave packet is constructed around a $\tilde p$ value
at which $\left . \partial^2 \varepsilon/\partial p^2\right  |_{\tilde{p}}=0$.
Although this is not possible for a free particle ($V(x)=0$) with 
$\varepsilon=p^2/2m$, it arises in the case a non-zero potential $V(x)$.
%The consideration of a possible point in the dispersion relation in which $\partial^2 \varepsilon/\partial p^2=0$ is 
%straightforward in this context, but usually not considered. 
In fact, in this case, the dispersion relation $\varepsilon(p)$ is periodic with periodicity $2\pi\hbar/a$,
and this implies that there are at least two points where $\partial^2 \varepsilon/\partial p^2=0$ (see Figure \ref{inflection}).
For the particular case of the tight binding model\cite{Kurt} with all-equivalent atoms, the dispersion relation becomes
\begin{equation}
\varepsilon (p)=-2W\cos(pa/\hbar),  
\end{equation}
with $W$ being an energy parameter associated with the jump probability of a particle between adjacent sites, which is taken as a constant.
We will analyze the spreading of a wave packet in such a situation.

\begin{figure}[h]
\includegraphics[width=14cm,clip=true]{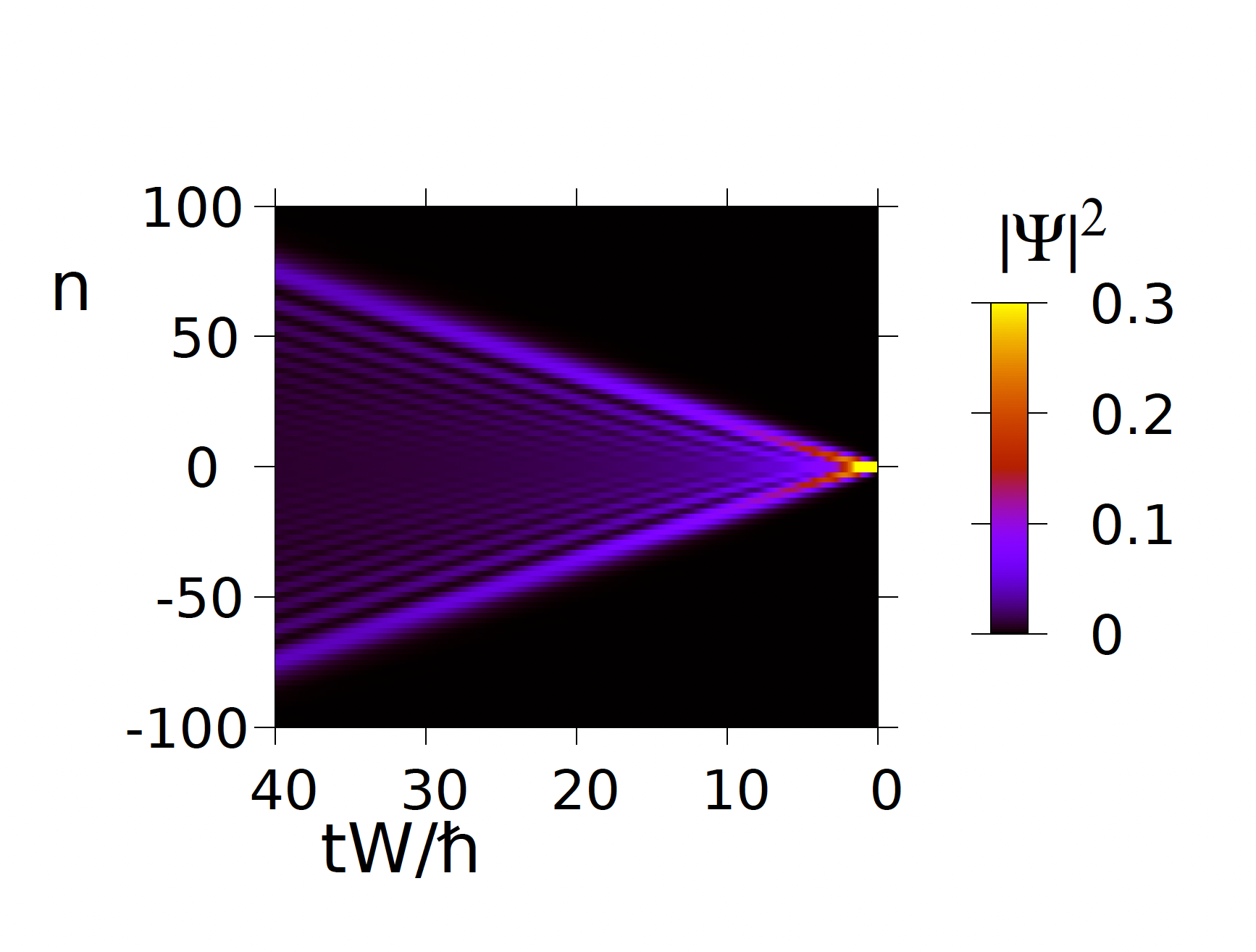}
\caption{(Color online) Space-time probability map ($|\Psi(n,t)|^2$) (where $n$ represents space, $x=na$) for a particle located at site 0, at $t=0$, of a one-dimensional tight-binding 
lattice. Compare with the result for surface perturbations in boat wakes (Fig. \ref{kelvin_completo}). Note that 
the additional oscillations seen in Fig. \ref{kelvin_completo} originate in the cosine factor in 
Eq. \ref{airycos} and do not appear in the present case.
}
\label{tight_binding}
\end{figure}

Consider the particular case of a particle that is in the site 0 at the initial time $t=0$.
The wave function of this particle at a time $t>0$ is obtained directly from (\ref{schro}), as
\begin{equation}
\Psi(n,t)=\frac{a}{2\pi \hbar}\int_{-\pi\hbar/a}^{+\pi\hbar/a} dp e^{{i\over \hbar}\left[pna+Wt\cos(pa/\hbar)    \right]},
\label{schro}
\end{equation}
where  we have set all $A_p=a/2\pi\hbar$ to fit the initial condition 
$\Psi(n,t=0)=\delta_{n,0}$.

Eq. (\ref{schro}) is an integral similar to Eq. (\ref{AiryCentral}) for the rainbow and Eq. (\ref{h}) for the boat wake. Specifically, the function $\epsilon (p)$, expanded around $p_0=\pm \pi \hbar/2a$, is of the form $\epsilon (p)\simeq  \mp 2W  \left[ (p-p_0) + {1\over 6} (p-p_0)^3\right]$, which is analogous to the cubic wavefront introduced in section II. 
Fig. \ref{tight_binding} shows the numerical solution of this equation. We clearly observe a maximum velocity in both directions, corresponding to the points of maximum group velocity, occurring at $p=p_0$, at which
$\partial^2 \varepsilon/\partial p^2=0$. These are the caustic lines of this problem: $n\simeq  \pm 2W t/\hbar$, and correspond to the Kelvin cone in the boat wake problem. 
The analogy extends further: in the quantum case, the ``wake" becomes a cone in space-time, with the caustic line tracing its boundary at $n/t=\pm 2W/\hbar$. The wave packet, initially localized at the origin, spreads due to a uniform jump probability, propagating in both directions along the one-dimensional chain. Waves corresponding
to other values of $p$ have lower velocity and interfere to form the Airy pattern   near the caustic lines—closely resembling the interference seen in phenomena such as rainbows and Kelvin wakes.
The  Airy pattern can be analytically worked out along the same lines as in  the previous sections.
The result is 

\begin{equation}
\Psi(n,t)=2\pi\left ({\frac {\hbar}{t W}}\right )^{1/3}\exp{(i\pi n/2)}   {\rm{Ai}}[(n-2W t/\hbar)/(tW/\hbar)^{1/3}].
\label{schro3}
\end{equation}
This is a wave packet that is only slightly dispersive. In fact,  its width increases as
$t^{1/3}, $\footnote{The value of the Airy function depends upon the combination $z = (n - 2Wt/\hbar)/(tW/\hbar)^{1/3})$. This means 
that, at a given time, the wave packet is shifted by an amount $2Wt/\hbar$ and stretched by a factor
$(tW/\hbar)^{1/3}$.} which is a slower rate than the typical increase as $\sim t^{1/2}$  when $\partial^2 \varepsilon/\partial p^2\ne 0$\cite{CohenT}. To our knowledge, this “space-time wake” has not been observed experimentally, but it may be observable in quantum cold atom systems.

%\textcolor{blue}{BTW, has this been observed in 1D quantum systems?}

%This result is obtained by noticing that the argument of the Airy function in Eq. (\ref{schro3}) is proportional
%to $\sim u/t^{1/3}$, where $u$ is the distance to the caustic line. Therefore, scaling time to $\lambda t$ increases the width of the packet in a %factor $\lambda^{1/3}$

%\textcolor{blue}{
%OTRA OPCION:     Note that in terms of   the scaled variable
%\[
%z = \frac{n - 2Wt/\hbar}{(tW/\hbar)^{1/3}},
%\]
%the Airy function $\rm{Ai}(z)$ is ``fixed". This means that, as a function of $n$, the wave packet is displaced by an amount $2Wt/\hbar$ and %``stretched" by an amount 
%$(tW/\hbar)^{1/3}$

%of a ship with a series of diagonal or oblique crests moving outwardly from the point of disturbance. These wave were first studied by Lord Kelvin 
%(http://en.wikip edia.org/wiki/Lord_Kelvin)

\section{Conclusions}
This paper has delved into the remarkable similarities between seemingly disparate wave phenomena: rainbows, boat wakes, and quantum mechanical wave packets. By emphasizing the underlying principles of ray folding, caustics, and Airy interference, we have highlighted the unifying concepts that bridge these diverse phenomena, as highlighted in Table \ref{airy_comparison}.

{\footnotesize
\begin{table}[h]
\centering
\begin{tabular}{|p{2.8cm}|p{6.8cm}|p{4cm}|}
\hline
\textbf{Phenomenon} & \textbf{Function involved} & \textbf{Coordinates } \\
\hline
Rainbow & 
\[
 e^{ik \rho} {\rm{Ai}}\left(-(kd)^{2/3}\alpha\right )
\] & 
$\alpha $ is the angle measured from the main rainbow. Fold caustic at the rainbow angle $\alpha=0$. \\
\hline
Boat Wakes & 
\[
{\tiny \cos \left(  \sqrt{\frac 32} x+\frac{\sqrt 3}{2}y  \right)
\times  {\rm{Ai}} \left[\frac{x+\sqrt 8 y}{(3{\sqrt{8}}y)^{1/3}}\right]}
\] & 
$x$ and $y$ are the distances from the boat, located at $(0,0)$. The caustic is at the Kelvin cone $x=-\sqrt{8}y$.\\
\hline
Quantum Packets & 
\[
{\rm{Ai}}\left[\frac{n-2W t/\hbar}{(tW/\hbar)^{1/3}}\right]
\] & 
The caustic appears at $n=2Wt/\hbar$ defining a ``Kelvin cone" in space-time.  \\
\hline
\end{tabular}
\caption{Comparison of  amplitude modulation given by the Airy function in the three phenomena described in this paper.}
\label{airy_comparison}
\end{table}
}

Through this comparative study, we hope to have provided a deeper appreciation of the interconnectedness of physical principles. By demonstrating how concepts from one domain can elucidate phenomena in another, we aim to enrich the pedagogical value of these topics and inspire further exploration into the beautiful and complex patterns observed in the natural world.

%\section{Acknowledgements}

\section{Appendix I. The Stationary Phase Approximation}

Even though the stationary phase method, also called the ``saddle point approximation" is commonly discussed in textbooks, we provide a brief overview here to ensure that our presentation is self-contained.

The approximation is relevant in cases where we have integrals of the form

$$
I=\int_{-\infty}^{\infty} dx e^{i f(x)},
$$
and $f(x)$ is a rapidly varying function of $x$. Rapid variations of $f(x)$ over most of the integration range mean that the integral averages almost to zero, except close to values of $x$ where $f(x)$ has an extremum. The approximation consists of expanding $f(x)$ around an extremum $x_0$, and evaluating the integral close to $x_0$.

The Taylor expansion of $f(x)$ around $x_0$, where $f'(x_0)=0$ is

$$
f(x)\simeq f(x_0) + {1\over 2} f'' (x_0) (x-x_0) ^2,
$$
which, substituted in the above integral gives

$$
I\simeq e^{i f(x_0)}\int_{-\infty}^{\infty} dx e^{i f''(x_0) (x-x_0)^2}= e^{i f(x_0)} \sqrt{\pi\over i f''(x_0)}.
$$
{Notice that the integral can be evaluated in the range $(-\infty, \infty)$ using the quadratic expansion of $f$ because the contribution from large values of $x-x_0$ to the integral is negligible.}

\subsection*{Application to the Airy integral}
%In the following appendix we apply this method to the specific case of the Airy function.

%\section{Appendix II. Airy, Young, and the Gouy's Phase}

%The success of Airy's theory of the rainbow has relegated Young's treatment to a secondary position. In fact, Airy himself described Young's approach as ``the imperfect theory" in his original presentation\cite{Airy}. Nonetheless, it is of interest to present some results that vindicate Young's treatment (see Ref. \cite{Laven} for a comprehensive discussion on this topic).

%We aim to compare Airy's and Young's treatments, particularly avoiding the caustic edge where Young predicts a nonphysical divergence in the intensity. Young's expression (Eq. (\ref{Young_completa})) is already sufficiently simple for any value of $\alpha$. To derive a similarly simple result from Airy's theory in this limit, we evaluate Airy's integral using the stationary phase method. We mention in passing that one of the first applications of the stationary phase approximation, and to our knowledge the first application to optics was Stokes evaluation of Airy's integral\cite{Stokes}.

%\textcolor{blue}{ No habría que ajustar esto con lo que está en el nuevo apéndice I ahora? O dicho de otra forma. El apéndice I está contenido acá !!
%YO EN CAMbio DIRIA en el Apendice A I que en el Apendice II mostramos un ejemplo concreto. }
For $x\gg 1$ the integral

\begin{eqnarray}
 {\rm{Ai}}(x)&=&\frac 1 {\pi}\int_{0}^{\infty}du ~{\cos(ux-u^3/3)}=\frac 1 {2\pi}\int_{-\infty}^{+\infty}du ~e^{i(ux-u^3/3)} 
\end{eqnarray}
has a rapidly varying argument of the exponential which has two extrema at $u=\pm \sqrt{x}\equiv \pm u_0$.
%In order to find the asymptotic ($x\gg 1)$ expression  for the  Airy integral, %note that the phase $vx-v^3/3$ is extremum for $v=\pm \sqrt{x}\equiv \pm v_0$. 
Expanding the phase to quadratic order around these extrema, we obtain:

\begin{eqnarray}
 2\pi{\rm{Ai}}(x)&=&\int_{-\infty}^{+\infty}du ~e^{i(ux-u^3/3)}\nonumber 
 \\
 &\simeq& e^{-i|u_0^{3}|}   \int_{-\infty}^{+\infty}du ~e^{iu_0(u-u_0)^2}+
 e^{i|u_0^{3}|}   \int_{-\infty}^{+\infty}du ~e^{iu_0(u+u_0)^2},
\end{eqnarray}

%The obtained Gaussian integral can be evaluated easily, 
and the results is
\begin{eqnarray}
 2\pi{\rm {Ai(x)}}\simeq e^{i|u_0|^{3}}   \sqrt\frac{\pi}{{i|u_0|}}+e^{-i|u_0|^{3}}   \sqrt\frac{\pi}{{-i|u_0|}}   \\
=\pi^{-1/2}x^{-1/4}\cos\left (\frac 23 x ^{3/2}-\frac \pi 4 \right )
\end{eqnarray}

Applying this result to Eq. (\ref{AiryCentral}), noting that $x=\alpha (kd)^{1/3}$ we obtain

\begin{equation}
U(\rho, \alpha) \simeq  {1\over \alpha ^{1/4}} {1\over \pi^{1/2} (kd)^{2/3}}\cos\left (\frac 23 (kd)\alpha  ^{3/2}-\frac \pi 4 \right ).
\end{equation}

%agrees 
%with Young's result modulo the 

 %\begin{equation}
 %    I\simeq
 %\end{equation}
 %\textcolor{blue}{deberia ser la expresion de Young con un lambda/4 corrido, no?}
 %which 
 The intensity $I\sim U^2$ coincides with Young's expression (Eq. (\ref{Young_completa2}))
 for $kd=({2\pi/\lambda})[R \sqrt{2/\theta''(x_0)}]$
  with an extra $\pi/4$ in the phase. This shift, the so-called Gouy's phase, is a subtle diffraction effect that occurs close to singularities such as focal points or  caustics (see Supplementary Material).  
 %In the next Appendix we discuss the essentials of this phase shift.

%\sout{As in the calculation in the rainbow one of the two emergent parallel rays passes through a 
%caustic while the other does not, the Young calculation should consider this $\pi/4$ difference, making 
%the final result much closer to the Airy one.}

\section{Supplementarry material to ``Common principles behind rainbows and boat wakes"}

\section*{Gouy's phase distilled to its essence}

%\textcolor{blue}{I'll add that the text of this section should become a separate supplementary file that will be shared online without further editing. But the instructions shared by Claire tell you how to write the text in order for readers to access the material.}

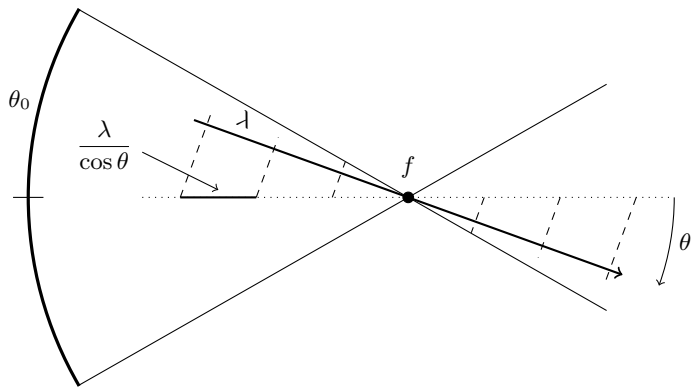
\begin{figure}
\begin{tikzpicture}[scale=1.]
%\draw [thick, ->](0.,0.)--(20.:3);
\draw [thick](0.,0.)--(160:3);
%\draw [thick] (0.,0.)--(-160:3);
\draw [thick,->](0.,0.)--(-20:3);
\draw[dotted] (-3.5,0)--(3.5,0);
\draw[dashed, shift={(-1,0)}] (0,0) -- (70:.5);
\draw[dashed, shift={(-2,0)}] (0,0) -- (70:1-.15);
\draw[dashed, shift={(-3,0)}] (0,0) -- (70:1.5-.33);
%
%\draw[dashed, shift={(-1,0)}] (0,0) -- (-70:.5);
%\draw[dashed, shift={(-2,0)}] (0,0) -- (-70:1-.15);
%\draw[dashed, shift={(-3,0)}] (0,0) -- (-70:1.5-.33);
%
\draw[dashed, shift={(1,0)}] (0,0) -- (-110:.5);
 \draw[dashed, shift={(2,0)}] (0,0) -- (-110:1-.15);
\draw[dashed, shift={(3,0)}] (0,0) -- (-110:1.5-.33);
%
%\draw[dashed, shift={(1,0)}] (0,0) -- (110:.5);
% \draw[dashed, shift={(2,0)}] (0,0) -- (110:1-.15);
%\draw[dashed, shift={(3,0)}] (0,0) -- (110:1.5-.33);
%
\node [rotate=-20] at (-2.15,1.05) {$\lambda$};
\draw[thick] (-3,0)--(-2,0);
\node  at (-4.,.7) {$\displaystyle{\lambda\over \cos\theta}$};
\draw [->](-3.5,.6) -- (-2.5,.1);
% \draw (20:3.3)--(20:3.7);
 \draw [->] (3.5,0) arc[start angle=0, end angle=-19.5, radius=3.5]
 node [midway,right] {$\theta$};
 \draw [fill] (0,0) circle (.07);
 \node at (0,.4) {$f$};
 \draw [very thick] (-5,0) arc[start angle=180, end angle=150, radius=5]
  node [midway,left] {$\theta_0$} ;
  \draw (-4.35,-2.5)--(.6*4.35,.6*2.5);
   \draw (-4.35,2.5)--(.6*4.35,-.6*2.5);
  \draw (-5.2,0)--(-4.8,0);
  \draw [very thick] (-5,0) arc[start angle=180, end angle=210, radius=5];
\end{tikzpicture}
\label{fig_gouy}
\caption{Plane waves of wavelength $\lambda$ are emitted from a circular arc, converging to a focal point $f$. Any wave originated at an angle $\theta\ne 0$ has an effective wavelength--when observed along the horizontal--larger than $\lambda$. This wavelength stretching  produces the Gouy's phase of the beam when passing through $f$.}
%\textcolor{blue}{Te parece poner dos lineas delgadas, que partan de los extremos del arco y pasen a través del foco hasta el otro lado?}}
\end{figure}

Let us consider  plane waves of wavelength $\lambda$ that originate from a coherent source which is an arc of circle spanning an angle of $2\theta_0$ (Fig. \ref{fig_gouy}).  On moving from left to right, the waves are convergent up to the focal point $f$ (where they arrive in phase), and then become divergent.
Let us evaluate the wave amplitude along the horizontal axis. Each wave forming a particular angle $\theta$ with the horizontal axis will produce along this axis a perturbation of an effective wavelength $\lambda_{eff}(\theta)\equiv\lambda/\cos(\theta)$, which is larger than $\lambda$. When summing up the contributions for all values of $\theta$ this ``stretching" of wavelength produces the Gouy's phase.

%Any of such waves with a particular value of $\theta$
%Let us concentrate on two plane waves forming angles $\pm \theta$ with the horizontal axis $x$.  Notice that the resultant wave along the axis of symmetry of the source (the horizontal line), is ``stretched" away from the focus \textcolor{blue}{La verdad que ahora no veo clara la explicación}, and has a wavelength $\lambda/\cos \theta$. 

To calculate the effect, we first notice that
the contribution of a wave with a particular value of $\theta$ to the amplitude $A$ along the $x$ axis is
\begin{equation}
 A(x,\theta)=A_0\cos(2\pi x/\lambda_{eff}(\theta))=   A_0\cos(kx\cos(\theta))
\end{equation}
with $k=2\pi/\lambda$, and where the coordinate $x$ is measured from the focal point.
We expand for small angles:
\begin{eqnarray}
\cos \left(kx\cos\theta\right)&\simeq& 
 \cos \left(k\left[1-{\theta^2\over 2}\right] x\right)\nonumber
\\ &=&
\cos (kx) \cos \left({k\theta^2\over 2}x\right) +\sin (kx) \sin \left({k\theta^2\over 2}x\right),
\end{eqnarray}
and  integrate over $\theta$ (using the additional approximation $kx\gg\theta_0$):

\begin{subequations}\label{Gouy}
	\begin{align}
		\int_{\theta_0}^{\theta_0}d\theta 
  \cos \left({k\theta^2\over 2}x\right)
  &\simeq 	\int_{\infty}^{\infty}d\theta 
  \cos \left({k\theta^2\over 2}x\right)
  =\sqrt{2\pi \over k |x|}
  \\
  \int_{\theta_0}^{\theta_0}d\theta 
  \sin \left({k\theta^2\over 2}x\right)
  &\simeq 	\int_{\infty}^{\infty}d\theta 
  \sin \left({k\theta^2\over 2}x\right)
  =\sqrt{2\pi \over k |x|} {\rm{sgn}}(x)
	\end{align}.
\end{subequations}

Then using elementary trigonometrical identities, we obtain the final result

\begin{eqnarray}
A(x)=A_0\int_{\infty}^{\infty}d\theta \cos \left(\left[k\cos\theta\right] x\right)&\simeq& 
2A_0\sqrt{2\pi \over k |x|} \left[
\cos(kx) + |\sin(kx)|\right] \nonumber \\
&=& 
4A_0\sqrt{\pi \over k |x|}
\cos\left( kx -{\rm{sgn}}(x) {\pi\over 4}
\right)
\end{eqnarray}

In other words, compared to the $\sim\cos (kx)$ amplitude oa a single wake that originates in the center of the arc, the nodes of the bundle of waves from a finite angular range are shifted by $\pi/4$ before reaching the focus, and by $-\pi/4$ after passing through it. 
The precise value $\pi/4$  results from the symmetry of the Fresnel integrals (Eq. (\ref{Gouy})) involved.
The total shift of $\pi/2$ occurs for all waves when passing through a focal point or a caustic, and is the Gouy's phase.

\end{document}